\documentclass[submission,copyright,creativecommons]{eptcs}
\providecommand{\event}{TTC 2013} 

\usepackage{multirow}
\usepackage{listings}
\usepackage{graphicx}
\usepackage{wrapfig}
\usepackage{multicol}
\usepackage{color}

\lstdefinelanguage{metaDepth}
{
keywordstyle=\color{black}\bfseries,
morekeywords={Node, Model, Edge, concept, bind, template, requires, strict,
              String, int, boolean, double, Date, self, forAll,
              abstract, imports, ordered, unique, extends,
              allInstances, size, and, operation, Integer, Set, load, id},
sensitive=true,
morecomment=[l]{--},
morestring=[b]',
showstringspaces=false
}

\lstdefinelanguage{textual}
{
keywordstyle=\color{black}\bfseries,
morekeywords={Node, Model, Edge, node, model, edge, template, template@1, template@2, for, Syntax,
              with, id, lingexts, flingexts, linginst, flinginst, typename, type, supers, override, imports},
sensitive=true,
morecomment=[l]{--},
morestring=[b]',
showstringspaces=false
}

\lstdefinelanguage{familyLanguage}{
        morekeywords={model, class, ref, scripting, emit, segment, local, queue, model, navigation, smap, attribution, mappings, patterns, tao, spat, composite, converter, convert, syn, inh, and, or, linking, with, chain, eclectic},
        morekeywords={from,to,end,uses,template,match,pattern, not, providing,!,<-,->, forall, exists},
        morekeywords={rule,module,from,to,create,uses,lazy,unique,helper},
        morekeywords={context,def,let,in,if,then,else,endif},
        morekeywords={and,or},
        morekeywords={do},
        emph={Boolean,Integer,String,Sequence},
        sensitive=true,
        morecomment=[l]{//},
        morestring=[b]',
        showstringspaces=false,
}

\lstdefinelanguage{binding}{
        morekeywords={binding,class,feature,is,to},
        morekeywords={helper,let,in},
        morekeywords={self,collect,if,then,else,endif},
        emph={Boolean,Integer,String,Sequence},
        sensitive=true,
        morecomment=[l]{--},
        morestring=[b]',
        showstringspaces=false,
}

\lstdefinestyle{nonumbers}{
	numbers=none,
}

\newcommand{\code}[1] {{\footnotesize\sffamily #1}}

\title{Solving the Flowgraphs Case with Eclectic}
\author{Jes\'us S\'anchez Cuadrado
\institute{Universidad Aut\'onoma de Madrid (Spain)}
\email{\quad jesus.sanchez.cuadrado@uam.es}
}
\def\titlerunning{Solving the Flowgraphs Case with Eclectic}
\def\authorrunning{Jes\'us S\'anchez Cuadrado}
\begin{document}
\maketitle

\begin{abstract}
This paper presents a solution for the
Flow Graphs case of the Transformation Tool Contest 2013, 
using the Eclectic model transformation tool.
The solution makes use of several languages of Eclectic,
showing how it is possible to combine them to address
a non-trivial transformation problem in a concise and
modulary way.
\end{abstract}

\section{Introduction}\label{sec:introduction}

The TTC 2013 Flow Graphs case~\cite{Horn13} proposes 
the analysis of Java programs,
conforming to the JaMoPP meta-model~\cite{jamopp}, by transforming
them into a language-independent meta-model which represents the
structure of the program and includes information about control
and data flows. 
This solution makes use of the Eclectic transformation
tool~\cite{eclectic} to solve the four proposed tasks. The fourth task has been addressed using
\textsc{MetaDepth}~\cite{metadepth} to create a small DSL.
\footnote{This solution is available as a SHARE image: \url{http://is.ieis.tue.nl/staff/pvgorp/share/?page=ConfigureNewSession&vdi=Ubuntu12LTS_TTC2013_Eclectic_FlowGraphs.vdi}}




Eclectic is a transformation tool designed as a family of model
transformation languages, that is, a set of transformation languages
each one specifically designed to address a specific transformation
concern, as well as some composition mechanisms for their
combination.  The objective of this solution is thus to show how it is
possible to address a non-trivial transformation task, such as this
case, using several languages and how this approach has the potential
of improving modularity and readability.

Eclectic currently provides the following languages: {\em i)} a
mapping language for establishing one-to-one and one-to-many
correspondences, {\em ii)} a target-oriented language with object
notation and explicit rule calls, {\em iii)} a traversal language
based on in the idea of attributed grammars, {\em iv)} a pattern
matching language which used object-notation, and {\em v)} a
lower-level scripting language, which also plays the role of
scheduling language.  Languages {\em i, ii and iv} do
not allow complex expressions, but these need to be encoded in
navigation libraries, written in the scripting language.

In principle, the combination of these languages permits covering many  
model transformation scenarios, in a more intentional way than using
a general purpose transformation language. Addressing case studies
could allow this intuition to be evaluated in practice.
The solution of this case has used the mapping language,
the attribution language, the pattern matching language, the scripting
language and a navigation module. The target-oriented language is
not needed because it is typically useful for synthesis tasks, but 
the case only involves mappings and analysis tasks.


In Eclectic every language is compiled to an
intermediate representation, called IDC. It provides primitive
instructions for model manipulation.  Then, IDC is compiled to the
Java Virtual Machine (JVM) bytecode format. In this way, all Eclectic
languages share the same execution infrastructure. The
composition mechanisms are implemented at the IDC level.
There is also a runtime library, which provides datatypes 
(e.g., immutable lists), a model manager
(i.e., EMF and \textsc{MetaDepth} are supported), etc.










\section{Solution}
\subsection{Task 1}\label{sec:task1}

The first task is a model-to-model transformation, which comprises
three different concerns that should be implemented in three different
modules:
i) A simple mapping between JaMoPP and FlowGraph elements must be
  performed. The mapping is mostly one-to-one, therefore the Eclectic
  mapping language would suffice. 
ii) A bottom-up text serialization of the JaMoPP abstract syntax 
  tree. This could be implemented with a series of helper methods
  or using the attribution language, which allows us to propagate
  text from the leaves of a statement to the root, creating the serialization
  during the process.
iii) An \code{Expression} element must not be translated, unless it
  is the condition of a {\em loop} or an {\em if}. To tackle this, 
  the pattern language would be in charge of recognizing the cases
  and it is combined with the mapping language.

In this way, the proposed solution makes use of three modules
(\code{task1\_map}, \code{task1\_attribution}, and \code{task1\_patterns}). 
The mapping module has a dependency on the attribution module,
to retrieve the textual representation of each source element,
and on the pattern matching module, which feeds it with 
non-trivial matches. 
The listing in Figure~\ref{fig:task1_mapping} shows an
excerpt of the transformation. It declares an Eclectic transformation
called \code{task1}, which encloses the three modules.

\begin{figure}[t!]
\begin{lstlisting}[language=familyLanguage, multicols=2]
eclectic task1 (in) -> (out)

mappings task1_map(in) -> (out) 
  uses task1_attribution  
  uses task1_patterns 

  from src :  in!ClassMethod
    to tgt : out!Method, exit : out!Exit
    linking tgt.exit = exit
      tgt.stmts <- src.statements
      tgt.txt  = task1_attribution!text[src]
  end	
	
  from src : in!WhileLoop
    to tgt : out!Loop
      tgt.expr <- src.condition
      tgt.body <- src.statement	
      tgt.txt  = task1_attribution!text[src]
  end

  from src : task_patterns!LoopExpression
    to tgt : out!Expr
      tgt.txt  = task1_attribution!text[src]	  
  end	  
end

attribution task1_attribution(in) -> (out)
  syn text : _!String

  rule in!WhileLoop
    text[self] <- "while"
  end
	
  rule in!AssignmentExpression
    left  = text[self.child]
    right = text[self.value]

    text[self] <- left.concat(' = ').concat(right)
  end
end

patterns task1_patterns(in)  
  def LoopExpression -> (e)
    l : in!WhileLoop {
      condition = e : in!Expression { }
    }
  end
  // ... Likewise for ConditionalExpression ...
end
\end{lstlisting}

\caption{Excerpt of the mapping from JaMoPP to FlowGraph}
\label{fig:task1_mapping}
\end{figure}

The mapping transformation is more or less straightforward.  Its
semantics is basically similar to ATL. Rules are executed at top level
(i.e., non-lazy execution), and 
the $\leftarrow$ operation (a binding) resolves a target element from a
source element. Interestingly, only simple expressions are allowed
in the right part of a binding. The most subtle detail is how to
``communicate'' with the other modules. 

To interoperate with the attribution transformation
the syntax \code{transformation!attribute[expr]} is used (see lines
11, 18 and 23), which means: {\em retrieve the element associated to
\code{expr} through the attribute}. As a concrete example, the text
for the \code{WhileLoop} (retrieved in line 18) is actually produced
by the assignment of the \code{text} attribute in line 31.

To interoperate with the pattern language, the mapping language treats
a pattern as a regular type. It can be seen as an extended layer put
on top of the original meta-model. In this way, the rule in lines
7--10 will be executed for each ocurrence of the \code{LoopExpression}
pattern, defined in lines 43--46. This pattern is matched if there is a
\code{WhileLoop} containing an \code{Expression} in this condition, 
and in such case the expression (variable \code{e}) is ``returned''.

The attribution transformation is also very simple, but the mechanics
of attributes has to be taken into account. The language supports
synthesized and inherited attributes (i.e., attributes propagated
bottom-up and top-down, respectively).\footnote{In practice, Eclectic
treats both types of attributes equally, but it is useful to
differentiate to improve readability.}

An attribute is assigned with the syntax \code{attribute[expr$_1$]
  $\leftarrow$ expr$_2$}, and it has the effect of creating a trace link
between the value obtained with \code{expr$_1$} and \code{expr$_2$}.
Conversely, retrieving the attribute associated to an element 
is done with the syntax \code{attribute[expr]}. For instance, 
in lines 35 and 36 the value of the text attribute is retrieved 
for the left and right parts of the assignment expression, and then
these two values are used to give the text value to the assignment expression,
that is, the {\em self} of the rule (line 38).

With respect to the integration at run-time of the different
modules, all modules are executed concurrently, exchanging data
among them as the execution proceeds. When all modules have finished
its execution, the transformation is finished.

\subsection{Task 2}\label{sec:task2}

This task is intended to complete the program structure
computed in the previous task with the links defining
the control flow graph of the program. 

It is an in-place transformation, as the source model has to
be augmented with the flow information. However, the main challenge
is the computation of the implicit flow relationships. This task
is particularly well suited for attribute-based traversal,
because control flow attributes have to be propagated along the
program structure (bottom-up and top-down).
The presented solution makes use of two attributes.
i) {\em successors} which is an inherited attribute specifying
   the list of ``flow'' siblings of each statement.  In addition, it
   relieves statements from knowning its position within its container
   statement.
ii) {\em cf\_next}, which is a synthesized attribute representing the
   flow instruction that corresponds to an element. This is useful
   to make the transformation more homogenous since every element will
   have a corresponding flow instruction (e.g., a Block)\footnote{A better name would be {\em cf\_instr}, since it does not represent the next control flow instruction (as cfNext does in the meta-model). The text, however, sticks to the name originally given in the solution uploaded to SHARE.}.


In this section only the rules for blocks and simple statements are shown
(see Figure~\ref{fig:task2_attr_blocks_simple_statements}),
just to give an impression
of the style of the solution. The complete explanation is given
in Appendix~\ref{app:task2}. 

The rule for \code{Block}\footnote{The type ends with ``!'' meaning that only instances of this type, but no subtypes, should be matched.} first
retrieves the block's \code{successors} (line 7) and propagates them
to the following sibling (line 8).  Then, it initializes the
attribute \code{successors}  for its enclosed statements (lines 11-13), adding
its first successor, so that the enclosed statements have an ``exit
point'' (i.e., this has the advantage that
there is no need to check if an element is the last one of a block).
Finally, the control flow instruction of a block is the
control flow instruction of the first enclosed statement (line 16), that
is, the flow reaches the block and goes on through the first statement.
Please note that for a series of nested blocks this approach 
will seamlessly work. The \code{cf\_next} attribute is thus used in the transformation
with the purpose of attaching a control flow instruction (a \code{FlowInstr}
element) to every element of the program tree, so that all elements can be homogenously treated as flow instructions even when
some of them are not \code{FlowInstr} elements, as it happens in the rule for \code{Block}

The rule for \code{SimpleStmt} first propagates the successors
to the immediate sibling (this operation has to be done in
every rule). Then,  it establishes that the flow instruction for 
the statement is
itself (line 23). Finally, the \code{cfNext} link is the control
flow instruction of its first successor.

\begin{figure}[h]
\begin{lstlisting}[language=familyLanguage, multicols=2]
attribution task2_attribution(flow) -> ()
  inh successors  : _!List
  syn cf_next   : flow!FlowInstr

  rule flow!Block!
    // Propagate the successors to immediate sibling
    successors = successors[self]
    successors[successors.first] <- successors.tail 

    // Initialize successors for the enclosed statements
    successor = successors.first
    successors[self.stmts.first] <- 
       self.stmts.tail.add(successor)
		
    // Compute the control flow
    cf_next[self] <- cf_next[self.stmts.first]		
  end

  rule flow!SimpleStmt
    successors = successors[self]
    successors[successors.first] <- successors.tail 

    cf_next[self] <- self
 
    next_flow = cf_next[successors.first]
    self.cfNext = next_flow
  end
\end{lstlisting}

\vspace{-10pt}
\caption{Computing the flow graph: blocks and simple statements}
\label{fig:task2_attr_blocks_simple_statements}
\end{figure}

\subsection{Task 3.1}\label{sec:task3_1}

This task complements Task 1 by adding variable declarations
to the FlowGraph models, and computing the information
about definitions and uses of the variables.

Thus, this transformation module (an attribution transformation)
depends on the mapping
transformation, so that its rules retrieve objects created by the latter.
To this end, the syntax \code{transformation!tlink.tfeature[expr]} is used,
which means: ``retrieve a trace link called {\em tlink} from
{\em transformation}, corresponding to the source element obtained
with {\em expr}''. A more detailed explanation about this feature
and the transformation itself is given in Appendix~\ref{app:task1_31}.

\subsection{Task 3.2}\label{sec:task3_2}

This task has been implemented using the straightforward algorithm
commented in the case description, using the scripting language. 
It was not possible to use attribute grammars because Eclectic does
not support circular dependencies yet.
Basically, for each variable 
use in a flow instruction, each path to
reach the instruction is looked up (using the \code{cfPrev} link). 
Then, for each path, every flow predecessor is computed
in a helper method ({\em all\_previous}). 
This works because {\em all\_previous} returns the list of
precedessors in order, so that if a variable is defined twice,
the closest predecessor is the first in the list.
The complete transformation is given in Appendix~\ref{app:task3_2}.

\subsection{Task 4}\label{sec:task_4}
This task requires building a small DSL to allow validation
specifications to be written. To this end 
the template language of \textsc{MetaDepth}~\cite{metadepth} has
been used.
It allows concrete syntaxes to be created ``on the fly'' (with
intermediate code generation, but it is handled internally).
\textsc{MetaDepth} is a powerful multi-level modeling framework, but 
its use here is very simple, so it is not fully introduced. 

The meta-model for the abstract syntax of the DSL is shown to the left
of Figure~\ref{fig:dsl_asm}. The model \code{ValidationDSL} acts as
root element, which encloses \code{RequiredLink} elements.  This
meta-class simply specifies that an instruction identified in
\code{left} must have the instruction identified in \code{right} as a
successor. The \code{ControlFlowLink} and \code{DataFlowLink} meta-classes
specialize \code{RequiredLink} for the control and data flow.

The right of Figure~\ref{fig:dsl_asm} shows the specification of the
concrete syntax. It is a template language, based on associating a
type with a specification of its serialization, which is later interpreted
to generate a parser. For instance, \code{\_:ControlFlowLinkTemplate}
invokes a template (line 7) and \code{\#left} (line 11) indicates
the serialization of the left property.

\begin{figure}[t!]
\begin{minipage}{.4\textwidth}
\begin{lstlisting}[language=metadepth]
Model ValidationDSL@1 {
  abstract Node RequiredLink {
     left  : String;
     right : String;
  }

  Node ControlFlowLink : RequiredLink 
  { }
  
  Node DataFlowLink : RequiredLink 
  { }
}
\end{lstlisting}
\end{minipage}\hfill
\begin{minipage}{.57\textwidth}
\begin{lstlisting}[language=textual]
load "validation_dsl"
Syntax ValidationDSLSyntax for ValidationDSL [".validate"] { 
   model template ValidationDSL@1 for "ValidationDSL" 
     "validate" ^Id 
   	(_:ControlFlowLinkTemplate)* (_:DataFlowLinkTemplate)* ;
      
   node template ControlFlowLinkTemplate@1 for ControlFlowLink
      "cfNext" ":" #left "-->" #right ;

   node template DataFlowLinkTemplate@1 for DataFlowLink
      "dfNext" ":" #left "-->" #right ;
}
\end{lstlisting}
\end{minipage}
\vspace{-10pt}
\caption{Meta-model of the DSL (left). Template specification (right)}
\label{fig:dsl_asm}
\end{figure}

The algorithm to check this specification against the generated models
basically consists of two nested loops, for traversing the
specification and the check model (see 
Appendix~\ref{app:task4}).

\vspace{-7pt}
\section{Evaluation}

\begin{wrapfigure}{r}{.35\textwidth}
\vspace{-35pt}
{\footnotesize
    \begin{tabular}{lll}
    Task        & Style                                                   & LOC            \\ \hline \hline
    1   & Mapping & 87\\ 
             & Attribute propagation & 160  \\ 
             & Simple pattern matching & 12  \\ \hline
    2   & Attribute propagation                         & 140             \\ \hline
    3.1 & Attribute propagation                                   & 123             \\ \hline
    3.2 & Scripting                                               & 40              \\\hline
    4   & Scripting & 102 \\ \hline
           & MetaDepth (meta-model)            & 10  \\
         & MetaDepth (c. syntax)            & 10  \\
   \hline
    Total   &                                  & 694 \\
          
   \hline 
   \end{tabular}
}
\end{wrapfigure}

All tasks have been solved, and the results for the smaller input models has been checked
manually. The only issue detected, in Task 3.2, has been missing data flow links 
for unary expressions.

With respect to comprehensibility and conciseness, the table  summarizes the use of the different languages of Eclectic and
the amount of code written (LOC, including whitespace). As has been
shown in the previous section, it was natural to combine different languages in order to
favour modularity, and ultimately readability through expressive and
concise specifications.

Finally, performance was not as good as expected. In particular, the control flow
transformation did not scale
well when large models were tried (notably tests 8 and 9). Therefore, a line of future work
is to profile and optimize the transformation engine.




{\footnotesize
\noindent {\bf Acknowledgements.} Work partially funded by the Spanish
Ministry of Economy and Competitivity (TIN2011-24139), and the R\&D
programme of Madrid Region (S2009/TIC-1650).
}

\vspace{-7pt}
\nocite{*}
\bibliographystyle{eptcs}
\bibliography{article}

\begin{thebibliography}{1}
\providecommand{\bibitemdeclare}[2]{}
\providecommand{\surnamestart}{}
\providecommand{\surnameend}{}
\providecommand{\urlprefix}{Available at }
\providecommand{\url}[1]{\texttt{#1}}
\providecommand{\href}[2]{\texttt{#2}}
\providecommand{\urlalt}[2]{\href{#1}{#2}}
\providecommand{\doi}[1]{doi:\urlalt{http://dx.doi.org/#1}{#1}}
\providecommand{\bibinfo}[2]{#2}

\bibitemdeclare{inproceedings}{eclectic}
\bibitem{eclectic}
\bibinfo{author}{Jes{\'u}s~S{\'a}nchez \surnamestart Cuadrado\surnameend}
  (\bibinfo{year}{2012}): \emph{\bibinfo{title}{Towards a Family of Model
  Transformation Languages}}.
\newblock {\sl \bibinfo{series}{LNCS}} \bibinfo{volume}{7307},
  \bibinfo{publisher}{Springer}, pp. \bibinfo{pages}{176--191},
  \doi{10.1007/978-3-642-30476-7\_12}.

\bibitemdeclare{techreport}{jamopp}
\bibitem{jamopp}
\bibinfo{author}{Florian \surnamestart Heidenreich\surnameend},
  \bibinfo{author}{Jendrik \surnamestart Johannes\surnameend},
  \bibinfo{author}{Mirko \surnamestart Seifert\surnameend} \&
  \bibinfo{author}{Christian \surnamestart Wende\surnameend}
  (\bibinfo{year}{2009}): \emph{\bibinfo{title}{{JaMoPP: The Java Model Parser
  and Printer}}}.
\newblock \bibinfo{type}{Technical Report} \bibinfo{number}{TUD-FI09-10},
  \bibinfo{institution}{Technische Universität Dresden, Fakult\"at
  Informatik}.
\newblock
  \bibinfo{note}{\url{ftp://ftp.inf.tu-dresden.de/pub/berichte/tud09-10.pdf}}.

\bibitemdeclare{inproceedings}{Horn13}
\bibitem{Horn13}
\bibinfo{author}{Tassilo \surnamestart Horn\surnameend} (\bibinfo{year}{2013}):
  \emph{\bibinfo{title}{The {TTC} 2013 Flowgraphs Case}}.
\newblock In: {\sl \bibinfo{booktitle}{Sixth Transformation Tool Contest (TTC
  2013)}}, {\sl \bibinfo{series}{EPTCS}} \bibinfo{volume}{this volume}.

\bibitemdeclare{inproceedings}{metadepth}
\bibitem{metadepth}
\bibinfo{author}{Juan \surnamestart de~Lara\surnameend} \&
  \bibinfo{author}{Esther \surnamestart Guerra\surnameend}
  (\bibinfo{year}{2010}): \emph{\bibinfo{title}{Deep Meta-Modelling with
  \textsc{MetaDepth}}}.
\newblock {\sl \bibinfo{series}{LNCS}} \bibinfo{volume}{6141},
  \bibinfo{publisher}{Springer}, pp. \bibinfo{pages}{1--20},
  \doi{10.1007/978-3-642-13953-6\_1}.

\end{thebibliography}

\appendix
\section{Complete code}

\subsection{Mapping to JaMoPP}\label{app:task1_31}
The following listing shows the code that solves Task 1 and Task 3.1. It is 
split into four modules. 
\begin{itemize}
\item A mapping module (\code{task1\_map}, lines 3--87).
\item An attribute computation module (\code{task1\_attribution}, lines 89--249).
\item An pattern matching module (\code{task\_patterns}, lines 258--263)  
\item An attribute computation module (\code{task3\_1\_varuses}, lines 269--391) 
\end{itemize}

As an implementation note, the expression language of Eclectic is
currently very simple, for instance, it does not have binary
expressions or if statements. The reasons is that it has not been
decided yet which style to use: a conventional one or a Smalltalk-like
(i.e., based on keyword methods).  In any case, by using method calls and
closures it is possible to express complex structures in practice
(although not in a very readable manner, see for example
lines 34--41 in Figure~\ref{fig:task2_attr_loops_and_conditionals}).
 

\begin{lstlisting}[language=familyLanguage, multicols=2]
eclectic task1 (in) -> (out)

mappings task1_map(in) -> (out)
   uses task1_attribution  as task1_attribution 
   uses task_patterns as task_patterns

   from src :  in!ClassMethod
     to tgt : out!Method, exit : out!Exit
     linking tgt.exit = exit
       tgt.stmts <- src.statements
			
       tgt.txt  = task1_attribution!text[src]
       exit.txt = "Exit"
   end	
	
   // -------------
   // Statements
   // -------------
	
   from src : in!LocalVariableStatement
     to tgt : out!SimpleStmt
       tgt.txt  = task1_attribution!text[src]
   end
	
   from src : in!ExpressionStatement
     to tgt : out!SimpleStmt
       tgt.txt  = task1_attribution!text[src]
   end
	
   from src : in!WhileLoop
     to tgt : out!Loop
       tgt.expr <- src.condition
       tgt.body <- src.statement	

       tgt.txt  = task1_attribution!text[src]
   end
	
   from src : in!Condition
     to tgt : out!If
       tgt.txt  = task1_attribution!text[src]
	  
       tgt.expr <- src.condition
       tgt.then <- src.statement	
       tgt.^else <- src.elseStatement
   end

   from src : in!Return
     to tgt : out!Return
       tgt.txt  = task1_attribution!text[src]
   end

   from src : in!Break
     to tgt : out!Break
       tgt.txt  = task1_attribution!text[src]	  
   end

   from src : in!Continue
     to tgt : out!Continue
       tgt.txt  = task1_attribution!text[src]	  
   end

   from src : in!JumpLabel
     to tgt : out!Label
       tgt.stmt <- src.statement
       tgt.txt  = task1_attribution!text[src]	  
   end
	
   from src : in!Block
     to tgt : out!Block
       tgt.stmts <- src.statements	
       tgt.txt  = task1_attribution!text[src]		
   end

   // -------------
   // Expressions
   // -------------
   from src : task_patterns!ConditionalExpression
     to tgt : out!Expr
       tgt.txt  = task1_attribution!text[src]
   end

   from src : task_patterns!LoopExpression
     to tgt : out!Expr
       tgt.txt  = task1_attribution!text[src]	  
   end	
	
end

attribution task1_attribution(in) -> (out)
//	optimizations : enabled
	syn text : _!String
	
	rule in!Method
		text[self] <- self.name.concat('()')
	end
	
	rule in!LocalVariableStatement
		init_text  = text[self.variable.initialValue]
		type_ref   = text[self.variable.typeReference]
		
		text[self] <- type_ref.concat(' ').concat(
		              self.variable.name.concat(' = ').
		              concat(init_text)).concat(';')
	end

	rule in!ExpressionStatement
		init_text  = text[self.expression]

		text[self] <- init_text.concat(";")
	end

	rule in!AssignmentExpression
		left  = text[self.child]
		right = text[self.value]
		operator = text[self.assignmentOperator]
		
		text[self] <- left.concat(' = ').concat(right)
	end

	rule in!SuffixUnaryModificationExpression
		expr_text = text[self.child]
		operator  = text[self.operator]
		text[self] <- expr_text.concat(operator) 
	end

	rule in!MultiplicativeExpression
		first = text[self.children.first]
		rest  = self.children.tail.zip(self.multiplicativeOperators)

		text[self] <- rest.inject(first) { |tmp, v| 
			tmp.concat(text[v.second]).concat(text[v.first])
		}
	end

	rule in!AdditiveExpression
		first = text[self.children.first]
		rest  = self.children.tail.zip(self.additiveOperators)
				
		text[self] <- rest.inject(first) { |tmp, v| 
			tmp.concat(text[v.second]).concat(text[v.first])
		}
	end

	rule in!RelationExpression
		first = text[self.children.first]
		rest  = self.children.tail.zip(self.relationOperators)
		
		text[self] <- rest.inject(first) { |tmp, v| 
			tmp.concat(text[v.second]).concat(text[v.first])
		}
	end

	rule in!EqualityExpression
		first = text[self.children.first]
		rest  = self.children.tail.zip(self.equalityOperators)		
		
		text[self] <- rest.inject(first) { |tmp, v| 
			tmp.concat(text[v.second]).concat(text[v.first])
		}
	end

	rule in!IdentifierReference
		text[self] <- self.target.name
	end

	rule in!DecimalIntegerLiteral
		text[self] <- self.decimalValue.to_s
	end

	rule in!WhileLoop
		text[self] <- "while"
	end

	rule in!Condition
		text[self] <- "if"
	end

	rule in!Block
		text[self] <- "{...}"
	end
	
	rule in!Continue
		text[self] <- "continue" 
	end
	
	rule in!Break
		text[self] <- "break"
	end

	rule in!Return
		rvalue = self.returnValue.is_nil.if_else({ 
			';' 
		}, { 
			v = text[self.returnValue]
			' '.concat(v.concat(';'))
		})
		text[self] <- "return".concat(rvalue)
	end

	rule in!JumpLabel
		text[self] <- self.name.concat(":")
	end

	// Types
	rule in!Int
		text[self] <- 'int'
	end
	
	// Operators
	rule in!Assignment
		text[self] <- ' = '
	end

	rule in!Multiplication
		text[self] <- ' * '
	end
	
	rule in!Addition
		text[self] <- ' + '
	end

	rule in!Division
		text[self] <- ' / '
	end
	
	rule in!Subtraction
		text[self] <- ' - '
	end

	rule in!Equal
		text[self] <- ' == '
	end	
	rule in!GreaterThan
		text[self] <- ' > '
	end

	rule in!LessThan
		text[self] <- ' < '
	end

	rule in!PlusPlus
		text[self] <- '++'
	end

	rule in!MinusMinus
		text[self] <- '--'
	end

end

patterns task_patterns(in)
	def LoopExpression -> (e)
		l : in!WhileLoop {
			condition = e : in!Expression { }
		}
	end

	def ConditionalExpression -> (e)
		l : in!Conditional {
			condition = e : in!Expression { }
		}
	end
end

// ----------------------
//       Task 3.1
// ----------------------

attribution task3_1_varuses(in) -> (out)
   uses task1_map as task1_map
    uses task_patterns as task_patterns

	inh vardef : out!Var
	syn writes : _!List
	syn reads  : _!List

	// Create variables
	rule in!Method
		translation = task1_map!default.t[self]
		vars = self.parameters.map { |p| 
			pvar = out!Param.new
			pvar.txt = p.name

			vardef[p] <- pvar
				
			pvar 
		}
		translation.vars = vars
		translation.^def = vars
	end
	
	rule in!LocalVariableStatement
		avar = out!Var.new
		avar.txt = self.variable.name

		vardef[self.variable] <- avar

		translation   = task1_map!default.t[self.up_to(in!Method)]
		translation.vars = avar
	end
	
	// Compute reads/writes for statements
	rule in!ExpressionStatement
		reads  = reads[self.expression]
		writes = writes[self.expression]

		translation = task1_map!default.t[self]
		translation.use  = reads
		translation.^def = writes
	end

	rule in!LocalVariableStatement
		left   = vardef[self.variable]
		reads  =  reads[self.variable.initialValue]
		writes = writes[self.variable.initialValue]
	
		translation = task1_map!default.t[self]
		translation.use  = reads
		translation.^def = writes.add(left)
	end

	rule in!Return
		self.returnValue.is_nil.if_false {
			reads = reads[self.returnValue]
			translation = task1_map!default.t[self]			
			translation.use = reads
		}
	end		
	
	rule in!UnaryModificationExpression
		avar = vardef[self.child.target]
		writes[self] <- avar.as_list
		reads[self]  <- avar.as_list	
	end
	
	// Compute reads/writes for expressions
	rule in!AssignmentExpression
		writes[self] <- vardef[self.child.target]
		reads[self]  <- reads[self.value]
	end	

	// covers ShiftExpression, AdditiveExpression, MultiplicativeExpression
	rule in!RelationExpression 
		writes[self] <- self.children.map { |c| r = writes[c] }.flatten 
		reads[self]  <- self.children.map { |c| r = reads[c] }.flatten 
	end		

	rule in!AdditiveExpression 
		writes[self] <- self.children.map { |c| r = writes[c] }.flatten 
		reads[self]  <- self.children.map { |c| r = reads[c] }.flatten 
	end		

	rule in!MultiplicativeExpression 
		writes[self] <- self.children.map { |c| r = writes[c] }.flatten 
		reads[self]  <- self.children.map { |c| r = reads[c] }.flatten
	end		

	rule in!EqualityExpression
		writes[self] <- self.children.map { |c| r = writes[c] }.flatten 
		reads[self]  <- self.children.map { |c| r = reads[c] }.flatten
	end		

	rule in!DecimalIntegerLiteral
		writes[self] <- _!List.new
		reads[self]  <- _!List.new
	end		
	
	rule in!IdentifierReference
		writes[self] <- _!List.new
		reads[self]  <- vardef[self.target]
	end		

	// Expressions
	rule task_patterns!LoopExpression
		translation = task1_map!default.t[self]
		reads  = reads[self]
		writes = writes[self]

		translation.use  = reads
		translation.^def = writes
	end	

	rule task_patterns!ConditionalExpression
		translation = task1_map!default.t[self]
		reads  = reads[self]
		writes = writes[self]

		translation.use  = reads
		translation.^def = writes
	end		
end
\end{lstlisting}


\subsection{Computing the control flow}\label{app:task2}


This transformation is perhaps the most complex one of the case,
so to simplify the explanation, the complete transformation has been
split into several listings. First, listing in
Figure~\ref{fig:task2_attr_methods_blocks} shows the header of the
transformation, including the attribute declarations (already
explained in Section ~\ref{sec:task2}),
 and the rules for \code{Method} and \code{Block}.


The rule for \code{Method}, initializes the \code{successors} attribute
for the first statement (line 7). It adds the \code{exit} element
to the list of sucessors as a fallback, so that the sucessor of the
last statement is the exit element (i.e., this has the advantage that
there is no need to check if an element is the last one of a block). 
Besides, the control flow instruction of \code{exit}
is itself. Lines 12--13 obtain the flow instruction for the first
statement, and set the \code{cfNext} link.

The rule for \code{Block} is similar to \code{Method}\footnote{The type ends with ``!'' meaning that only instances of this type, but no subtypes, should be matched.}, but first it
retrieves the block's \code{successors} (line 19) and propagates them
to the following sibling (line 20).  Then, it initializes the
\code{successors} attribute for its statements (lines 23-25), adding
its first successor, so that the enclosed statements have an ``exit
point''. Finally, the control flow instruction of a block, is the
control flow instruction of the first enclosed statement (line 28).
Please note that for a series of nested blocks this approach 
will seamlessly work.

\begin{figure}[h]
\begin{lstlisting}[language=familyLanguage, multicols=2]
attribution task2_attribution(flow) -> ()
  inh successors  : _!List
  syn cf_next   : flow!FlowInstr

  rule flow!Method
    // Initialize sucessors for enclosed stmts
    successors[self.stmts.first] <- 
       self.stmts.tail.add(self.exit)

    cf_next[self.exit] <- self.exit 
		 
    // Set flow link with the first flow instruction
    next_flow   = cf_next[self.stmts.first]
    self.cfNext = next_flow
  end
	
  rule flow!Block!
    // Propagate the successors to immediate sibling
    successors = successors[self]
    successors[successors.first] <- successors.tail 

    // Initialize sucessors for the enclosed statements
    successor = successors.first
    successors[self.stmts.first] <- 
       self.stmts.tail.add(successor)
		
    // Compute the control flow
    cf_next[self] <- cf_next[self.stmts.first]		
  end
\end{lstlisting}
\caption{Computing the flow graph: methods and blocks}
\label{fig:task2_attr_methods_blocks}
\end{figure}

Once the two basic enclosing structures have been presented,
the easiest elements are simple statements (\code{SimpleStmt})
and returns (\code{Return}), which are addressed in the
listing shown in Figure~\ref{fig:task2_attr_simple_and_return}.

The rule for \code{SimpleStmt} first propagates the successors
to the immediate sibling (this operation has to be done in
every rule, so it will not be explained in the following). Then, 
it establishes that the flow instruction for the statement is
itself (line 5). Finally, the \code{cfNext} link is the control
flow instruction of its first successor.

In contrast, the rule for \code{Return} needs to look up the \code{Method}
in which the instruction is enclosed, in order to set the \code{cfNext}
link to the method's exit element (lines 16--17). The \code{up\_to}
facility returns the first ancestor with the given type.

\begin{figure}[t!]
\begin{lstlisting}[language=familyLanguage, multicols=2]
rule flow!SimpleStmt
  successors = successors[self]
  successors[successors.first] <- successors.tail 

  cf_next[self] <- self
 
  next_flow = cf_next[successors.first]
  self.cfNext = next_flow
end
rule flow!Return
  successors = successors[self]
  successors[successors.first] <- successors.tail 
	
  cf_next[self] <- self

  method = self.up_to(flow!Method)
  self.cfNext = method.exit		
end
\end{lstlisting}
\caption{Computing the flow graph: simple statements and return}
\label{fig:task2_attr_simple_and_return}
\end{figure}

The approach for \emph{loops} and \emph{conditionals} follows
a similar schema, but taking into account that the actual flow
instruction is their condition, as well as 
the particularities of each instruction.
The solution is shown in the listing of Figure~\ref{fig:task2_attr_loops_and_conditionals}.

In the case of \code{Loop}, the \code{successors} attribute for
its body has to be the condition expression, that is, the
control flow successor of the loop's last statement will be
the loop's condition (lines 5--6). The control flow instruction
of the loop is its condition, and the control flow of the
condition is itself (this is needed because other instructions
will refer to the control flow instruction of the condition as
it has been designated the successor of the loop). Finally,
the \code{cfNext} link is set to the next successor as usual,
but also to the first enclosed flow instruction (lines 11--15).

The solution for conditionals (meta-class \code{If}, lines 22--42) is
conceptually easier. The successors of the \code{then} part are the
if's successors (line 26), the flow instruction is its condition (line
29) and the successor of the condition is the instruction within the
\code{then} (lines 31--32). Finally, it requires checking whether there is
an {\em else} part (line 34)\footnote{This syntax for conditionals is
  only a syntatic limitation, as the current expression language is
  kept to a minimum.}. If not, the next control flow instruction is just the
following successor (lines 35--36).  Otherwise, the successor
attribute has to be initialized for the else part, and the 
next control flow instruction is the one within the \code{then} part (lines 38--40).

\begin{figure}[t!]
\begin{lstlisting}[language=familyLanguage, multicols=2]
rule flow!Loop
  successors = successors[self]
  successors[successors.first] <- successors.tail 

  condition  = self.expr
  successors[self.body] <- condition.as_list 

  cf_next[self] <- condition		
  cf_next[condition] <- condition

  next_flow = cf_next[successors.first]
  condition.cfNext = next_flow

  first_within = cf_next[self.body]
  condition.cfNext = first_within					
end





rule flow!If
  successors = successors[self]
  successors[successors.first] <- successors.tail 

  successors[self.then] <- successors 
		
  condition  = self.expr
  cf_next[self] <- condition
  
  first_then = cf_next[self.then]
  condition.cfNext = first_then		

  self.else.is_nil.if_else({
    next_flow = cf_next[successors.first]
    condition.cfNext = next_flow				
  }, {
    successors[self.else] <- successors
    first_within = cf_next[self.else]
    condition.cfNext = first_within				
  })
end
\end{lstlisting}
\caption{Computing the flow graph: loops and conditionals}
\label{fig:task2_attr_loops_and_conditionals}
\end{figure}

Finally, rules to deal with \code{Break} and \code{Continue}
statements (including \code{Labels}) are introduced. In both
cases, the key issue is to determine the jump location, which
will be different depending on whether there is a label or not.
The listing in Figure~\ref{fig:task2_attr_break_and_continue} shows
the solution.

In the case of a \code{Break}, the jump location is the 
enclosing loop or the label (lines 8-12). Then, the next
flow instruction is simply the successor of the jump location
(lines 14--16).

In the case of a \code{Continue}, the jump location is assumed
to be the condition expression of a loop, either the enclosing
loop or a loop with a label assigned (lines 25--30). Thus,
the next flow instruction is just this expression (line 32).

Finally, for a \code{Label} the control flow instruction is the 
control flow instruction of the statement that it is labelling (line 41).

\begin{figure}[h]
\begin{lstlisting}[language=familyLanguage, multicols=2]

rule flow!Break
  successors = successors[self]
  successors[successors.first] <- successors.tail 

  cf_next[self] <- self
		
  jump_location = self.label.is_nil.if_else({
    self.up_to(flow!Loop) 
  }, {
    self.label
  })
		
  break_successors = successors[jump_location]
  next_flow = cf_next[break_successors.first]
  self.cfNext = next_flow			
end
	
rule flow!Continue
  successors = successors[self]
  successors[successors.first] <- successors.tail 

  cf_next[self] <- self

  expr = self.label.is_nil.if_else({
    loop = self.up_to(flow!Loop) 
    loop.expr	
  }, {
    self.label.stmt.expr
  })

  self.cfNext = expr
end
	
rule flow!Label
  successors = successors[self]
  successors[successors.first] <- successors.tail 
  successors[self.stmt] <- successors

  cf_next[self] <- cf_next[self.stmt] 
end
\end{lstlisting}
\caption{Computing the flow graph: break and continue}
\label{fig:task2_attr_break_and_continue}
\end{figure}


\subsection{Computing the data flow}\label{app:task3_2}
The listing in Figure~\ref{fig:task3_dataflow} shows the implementation
of this task. There is a navigation module \code{task3\_2\_navigation} which
adds the method \code{all\_previous} to \code{FlowInstr} elements, so that
it can be used by \code{task3\_2\_attribution} to set the data flow links.

\begin{figure}[t!]
\begin{lstlisting}[language=familyLanguage, multicols=2]
navigation task3_2_navigation(flow)

   def flow!FlowInstr.all_previous
     visited_map = _!Map.new.^put(self, true)
     self.all_previous_aux(visited_map)
   end

   def flow!FlowInstr.all_previous_aux(visited)
     not_visited = self.cfPrev.
        reject { |p| visited.include(p) }

     previous = not_visited.map { |p|
       p.all_previous_aux(visited.^put(p, true))
     }.flatten
		
     self.as_list.concat(previous.concat(not_visited))
   end

end


attribution task3_2_attribution(flow) -> ()
  uses task3_2_navigation

  rule flow!FlowInstr
    self.use.each { |v|
      // Look in each of the paths
      self.cfPrev.each { |i|
        def_instruction = i.all_previous.select { |prev| 
          prev.^def.include(v)
        }.first
				
        def_instruction.dfNext = self
      }			

      self.^def.include(v).if_true {
        self.dfNext = self
      }
    }
  end		
end

\end{lstlisting}
\caption{Computing the data flow}
\label{fig:task3_dataflow}
\end{figure}

It is worth mentioning that a solution based on attribute
propagation, following the algorithm proposed in the Dragon Book
was tried, but it requires circular attributes, which are
currently not supported in Eclectic.
Nevertheless, this solution shows that navigation modules are
also possible, as well as scripting-based transformations.

\subsection{Checking control and data flow models}\label{app:task4}
The comparison of the control of the data flow models against the
validation specification expressed with the DSL created in
Section~\ref{sec:task_4} has been implemented with the Eclectic low-level scripting
language. Interestingly, the Eclectic high-level languages are
compiled to a representation similar to this one, so this explanation
may serve to give the reader an intuition of how Eclectic works under
the hood.

The program shown in Listing~\ref{fig:task4_scripting} takes two input
models, the specification written with the DSL and the flow graph
model. It outputs a report model (actually, the current implementation
just prints the reports, but it will be straightforward to create
elements of the report model).

The scripting transformation allows temporary data structures to be
defined, which serve as intermediate data for the transformation.
In this way, lines 2--7 defines a model called \code{inline},
with the \code{FlowLink} class. This class will hold a control
flow or data flow relationship in the form of a string
representing the source element and another string representing
the target element.

Afterwards, queues are defined. In the scripting language
(and in IDC, the intermediate representation used by Eclectic)
communication happens through queues. A model queue (lines 9--11)
declares the interest of a transformation in a certain type.
A local queue (lines 13--17) is used internally 
by communicating values between two places of the transformation.
The \code{flow\_cfLinks} and \code{flow\_dfLinks} will contain
links appearing in the \code{flow} model, and the
\code{dsl\_cfLinks} and \code{dsl\_dfLinks} will contain links
appearing in the DSL specification.

The transformation code can be logically organised into
\code{segment}s. In this way, the \code{find\_flow\_links} segment
(line 19) contains code to find flow links. The \code{forall} 
instruction is able to receive elements of a queue (e.g., line 20).
The \code{emit} instruction sends an object to a queue, in particular
it is used to send \code{FlowLink} elements when a link is found (e.g., line 25).
This is the basic communication mechanism between patterns and rules
(although in this language the distinction is implicit).

Then, segment \code{validate} (lines 52--98) receives the notifications
of the found flow links (through the four local queues) and check
false links and missing links. As Eclectic has full support for closures,
it is possible to declare a closure as if it were a local variable,
acting as kind of local function. This is done, for example, in
lines 53--60 to create a facility to check false links.

\begin{figure}[t!]
\begin{lstlisting}[language=familyLanguage, multicols=2]
scripting task4_script(dsl, flow) -> (report)	
  model inline
    class FlowLink
      ref source : _!String
      ref target : _!String
    end
  end
	
  model queue mFlowInstr : flow!FlowInstr
  model queue mControlFlowLink : dsl!ControlFlowLink
  model queue mDataFlowLink    : dsl!DataFlowLink
	
  local queue flow_cfLinks : inline!FlowLink
  local queue flow_dfLinks : inline!FlowLink

  local queue dsl_cfLinks  : inline!FlowLink
  local queue dsl_dfLinks  : inline!FlowLink
	
  segment find_flow_links
    forall flow_instr from mFlowInstr
       flow_instr.cfNext.each { |target|
          lnk = inline!FlowLink.new			
          lnk.source = flow_instr.txt
          lnk.target = target.txt
          emit lnk to flow_cfLinks
       }

       flow_instr.dfNext.each { |target|
          lnk = inline!FlowLink.new			
          lnk.source = flow_instr.txt
          lnk.target = target.txt
          emit lnk to flow_dfLinks
       }			
     end

     forall control_flow from mControlFlowLink
       lnk = inline!FlowLink.new			
       lnk.source = control_flow.left
       lnk.target = control_flow.right
       emit lnk to dsl_cfLinks
     end
		
     forall data_flow from mDataFlowLink
       lnk = inline!FlowLink.new			
       lnk.source = data_flow.left
       lnk.target = data_flow.right
       emit lnk to dsl_dfLinks
     end
		
   end
	
   segment validate
     check_false_link = { |type, lnk, dsl_links|
       dsl_links.find { |cf| 
         cf.left.eq(lnk.source).and(
           cf.right.eq(lnk.target))
       }.if_nil {
          lnk.source.concat(' ==> ').concat(lnk.target).println(type.concat(" false link: "))
       }			
     }

     dsl_expected_cfs = dsl!ControlFlowLink.all_instances		
     dsl_expected_dfs = dsl!DataFlowLink.all_instances

     // For any cfNext or dfNext link in the model, 
     // check if it is also defined in the spec.
     forall cfLink from flow_cfLinks
       check_false_link.call('Control', cfLink, dsl_expected_cfs)
     end
		
     forall dfLink from flow_dfLinks
       check_false_link.call('Data', dfLink, dsl_expected_dfs)
     end

     // Check that every link in the specification 
     // occurs in the flow graph
     flow_instrs = flow!FlowInstr.all_instances		

     check_missing_link = { |type, lnk, featureName|
       flow_instrs.find { |fi| 
           next_txt = fi.get(featureName).map { |n| n.txt }
                               
           fi.txt.eq(lnk.source).
           and(next_txt.include(lnk.target))
       }.if_nil {
           lnk.source.concat(' ==> ').concat(lnk.target).println(type.concat(" missing link: "))
       }			
     }
      
     forall dsl_cfLink from dsl_cfLinks
       check_missing_link.call('Control', dsl_cfLink, 'cfNext')
     end
		
     forall dsl_dfLink from dsl_dfLinks
       check_missing_link.call('Data', dsl_dfLink, 'dfNext')
     end		
  end
end
\end{lstlisting}
\caption{Validating the flow graph using the scripting language}
\label{fig:task4_scripting}
\end{figure}

\end{document}